\documentclass[aps,prb,twocolumn,10pt,superscriptaddress,floatfix,a4paper]{revtex4-1}

\usepackage{amsmath}
\usepackage{amssymb}
\usepackage{mathrsfs}
\usepackage{graphicx}
\usepackage{titlesec}
\usepackage{titleref}

\renewcommand{\vec}{\textbf}

\begin{document}

\title{Facing the phase problem in Coherent Diffractive Imaging via Memetic Algorithms}
\author{Alessandro Colombo}
\email{alessandro.colombo6@unimi.it}
\affiliation{Universit\`a degli Studi di Milano, via Giovanni Celoria 16, 20133 Milano, Italy}

\author{Davide Emilio Galli}
\email{davide.galli@unimi.it }
\affiliation{Universit\`a degli Studi di Milano, via Giovanni Celoria 16, 20133 Milano, Italy}

\author{Liberato De Caro}
\email{liberato.decaro@ic.cnr.it}
\affiliation{Istituto di Cristallografia, Consiglio Nazionale delle Ricerche  (IC-CNR), via Giovanni Amendola, 122/O, 70126 Bari, Italy}

\author{Francesco Scattarella}
\email{scattarella@iom.cnr.it}
\affiliation{Istituto Officina dei Materiali, Consiglio Nazionale delle Ricerche (IOM-TASC-CNR), Strada Statale 14 km 163.5, 34149 Trieste, Italy }

\author{Elvio Carlino}
\email{carlino@iom.cnr.it}
\affiliation{Istituto Officina dei Materiali, Consiglio Nazionale delle Ricerche (IOM-TASC-CNR), Strada Statale 14 km 163.5, 34149 Trieste, Italy}

\begin{abstract}
Coherent Diffractive Imaging is a lensless technique that allows imaging of matter at a spatial resolution not limited by lens aberrations. 
This technique exploits the measured diffraction pattern of a coherent beam scattered by periodic and non--periodic objects to retrieve spatial 
information. The diffracted intensity, for weak--scattering objects, is proportional to the modulus of the Fourier Transform of the object scattering function. Any phase information, needed to retrieve its scattering function, has to be retrieved by means of suitable algorithms. 
Here we present a new approach, based on a memetic algorithm, i.e. a hybrid genetic algorithm, to face the phase problem, which exploits 
the synergy of deterministic and stochastic optimization methods. The new approach has been tested on simulated data and applied 
to the phasing of transmission electron microscopy coherent electron diffraction data of a $\text{SrTiO}_\text{3}$ sample. 
We have been able to quantitatively retrieve the projected atomic potential, and also image the oxygen columns, which are not directly visible 
in the relevant high-resolution transmission electron microscopy images. Our approach proves to be a new powerful tool for the study of matter 
at atomic resolution and opens new perspectives in those applications in which effective phase retrieval is necessary.
\end{abstract}

\maketitle

\section*{Introduction}
Full-field and scanning microscopes can be either lens-based or  lensless imaging systems. Coherent Diffractive Imaging (CDI) is a  lensless technique that  permits imaging matter at a spatial resolution not limited by lens aberrations. The seminal idea of CDI was due to David Sayre in 1952 \cite{Sayre1952} but it was only experimentally demonstrated for X-rays in 1999 \cite{Miao1999} and, more recently, also for electrons, using a Transmission Electron Microscope (TEM), giving rise to the Electron Diffractive Imaging (EDI) \cite{Zuo2003,Huang2009,DeCaro2010}. 

The goal is to retrieve a qualitative/quantitative image of a scattering function related to a physical property of the scattering object, such as the electron density (X-ray CDI) or the atomic potential (EDI).
High Resolution TEM (HRTEM) images of the projected atomic potential are phase-contrast images limited by the high-order aberrations of the objective lens, which distort the phase of the scattered wave function, giving rise to images of the sample, which in general are not immediately interpretable in terms of its atomic structure \cite{Spence1988}.
Instead, diffraction patterns of scattering objects are not affected by these aberrations. Therefore they contain, in principle, undistorted information on the scattering function at a better spatial resolution with respect to lens-based imaging systems \cite{DeCaro2010, Huang2009, Zuo2003}. The diffracted intensity, for weak-scattering objects, is proportional only to the modulus of the Fourier Transform (FT) of the scattering function. Any phase information, which is experimentally lost (phase problem \cite{Sayre1952}), has to be retrieved by means of suitable algorithms. The lensless image of the scattering function, obtained by means of an inverse FT of the diffraction pattern once that the correct phase has been retrieved, is characterized by a final resolution experimentally limited only by the Numerical Aperture ($\text{NA}_\text{diff}$) corresponding to the highest spatial frequency contained in the diffraction pattern that can be related to the atomic structure of the investigated sample \cite{DeCaro2010}. Wavelength, noise, radiation damage, thermal and mechanical stability of the experimental setup, dynamics of the detector, etc. \cite{Decaro2013} could limit the spatial resolution achievable. 

In order to find a unique solution to the phase problem, phase retrieval algorithms need a-priori constraints, such as fixing a region around the sample characterized by zero scattering \cite{Fienup1982}. The extension of the zero-scattering region has to be large enough regarding the object support \emph{S}, containing the scattering object. This requirement is the so-called oversampling of the diffraction pattern, needed to satisfy the Nyquist sampling requirements \cite{Sayre1952}.

For extended samples, alternative approaches have been developed to have enough a-priori information to make the phasing problem overdetermined. In particular, Abbey et al.   \cite{Abbey2008} for X-rays CDI experiments substituted the region of zero scattering with a region of zero illumination using a confined probe with a technique named keyhole CDI. The possibility to perform Keyhole coherent diffraction experiments has been demonstrated also for electrons in EDI, obtaining the KEDI approach \cite{DeCaro2012}.

So far, several phase retrieval algorithms have been developed \cite{shechtman2015}, which are mostly evolutions of Fienup’s modification \cite{Fienup1982} of the Gerchberg-Saxton's algorithm \cite{Gerchberg1972} working with a dual-space strategy. 
The common approach of many phasing algorithms is to impose constraints both in real space (the prior information on the zero scattering/illuminating region) and in Fourier space (the amplitudes are adapted to the experimental values). The imposition of a constraint in one space always causes the violation of the constraint in the other. Consequently, the standard strategy is to use iterative schemes but, in this way, the global minimum of the reconstruction errors is often reached with difficulty \cite{Marchesini2007}. Indeed, the support \emph{S} of the scattering function is either unknown (X-ray CDI) or known at a worse spatial resolution. This is the case of EDI, for example, where the support is obtained by means of a lens-based image of the scattering function (projected atomic potential), i.e. by the HRTEM image. 

In case of unknown complex scattering functions the knowledge of the support \emph{S} (non-zero illumination region) at the same spatial resolution of the measured diffraction pattern seems to be mandatory for the success of the phasing, especially when phasing algorithms are not suitably structured to escape stagnation in local minima \cite{DeCaro2010,DeCaro2012}. This aspect would limit the advantages of indirect imaging based on lensless systems with respect to conventional lens-based set-ups.

Indeed, standard phasing approaches, such as the Hybrid Input-Output (HIO), are mainly \emph{deterministic} iterative algorithms, which, going back and forth from the real to the Fourier space, try to optimize a specific error functional \cite{Fienup1982}. For this reason they are highly efficient in finding local minima, but they suffer  from stagnation mainly due to the incomplete knowledge of the support S. Furthermore, the final result is highly dependent on the initial conditions \cite{Marchesini2007}.
In order to overcome these limitations, phasing procedures usually involve a lot of parallel and independent retrieval processes, with different initial conditions, choosing the scattering function with the lowest reconstruction errors as a possible solution.
A first step toward a smarter use of information coming from multiple phase retrieval processes is the Guided Hybrid Input-Output algorithm \cite{chen2007}.

A different approach to face the phase problem in CDI could make use of pure stochastic optimization methods. However, a major limitation of these methods is their inefficency once the number of unknowns is huge, so that, in typical CDI applications, such an approach is doomed to fail even by exploiting actual super--computing facilities.

In this article we make a step further proposing a new hybrid stochastic approach to better explore the phase solution space through a smart use of Genetic Algorithms (GAs) \cite{Goldberg1989}. GAs have been already applied to the phase problem in different fields \cite{thust1997, nicholson1999, taylor2006, li2011}.
The novelty of our new approach consists in the development of a Memetic Algorithm (MA) \cite{ong2010} 
in the context of phase retrieval applied to CDI; this scheme represents a natural choice for a smart merging 
of stochastic and deterministic optimization methods: the algorithm has been developed hybridizing a GA, which guarantees a wide exploration of the configuration space, with local optimization algorithms like Hybrid Input-Output and Error Reduction.

We have shown on simulated data that the MA phasing approach is able to retrieve the correct scattering function, when it is real, imposing a very loose support constraint \emph{S} in the direct space. 
Moreover, still on simulated data, we have shown that the new MA phasing approach is able to retrieve the correct scattering function, even if it is complex, starting from a knowledge of the support \emph{S} at a resolution four times worse than the one corresponding to $\text{NA}_\text{diff}$, even worse than normally observed in EDI/KEDI real experiments. Our new approach shows convergence performances towards the global minimum that go well beyond those achievable by standard phasing \emph{deterministic} algorithms. 

Finally, we have applied the GA-based phasing approach to a KEDI experiment realized on a $\text{SrTiO}_\text{3}$ sample in a [100] axis orientation.
The image obtained after the phase reconstruction is a detailed structural map of the specimen atomic potential projected along the [100] direction  at a sub-\r{A}ngstr\"om spatial resolution corresponding to the highest frequency measured in the experimental diffraction pattern. The intensity distribution enables one to distinguish between atomic sites containing different chemical species. Also the oxygen signals can be detected, despite the presence of heavy atoms in the crystal cell, whereas they are not visible in the relevant HRTEM image.

These results pave the way to the highest spatial resolution, accuracy and reliability achievable in lensless imaging and represent a new powerful tool for the study of matter.

\section*{Results and discussion}

\subsection*{The Memetic Phase Retrieval approach }\label{MPR}

A GA \cite{Goldberg1989} is a stochastic optimization method that imitates the survival--to--fitness typical of the natural evolution of a population. In general, this is obtained by elaborating the genetic information via three genetic operators: \emph{Selection}, \emph{Crossover} and \emph{Mutation}.
In our algorithm we induce the genetic dynamics on a set of initial densities $ \{ \rho_i(\vec{x}) \}_{i=1 \ldots N_p}$ (also called \emph{population}) which represents $N_p$ possible solutions to the phase problem. 
In standard GAs \emph{Mutation} and \emph{Crossover} induce a stochastic shift on every element of the population in the space of configurations; this improves the ability of exploring the space, but makes the GA efficient only when $N_p$ increases with the number of unknowns \cite{goldberg1989sizing}.
In the phase retrieval problem this condition makes standard GA impracticable due to the computational cost.

A practised way to overcome this issue consists in implementing hybrid GAs, known as Memetic Algorithms (MAs) \cite{moscato1989,renders1996,el2006,Neri2012}, where global optimization is boosted by local optimization procedures. This operation, in the framework of Memetic Algorithms, is known as \emph{Self} (or \emph{Local}) \emph{Improvement}.

\begin{figure}
\centering
  \includegraphics[width=\linewidth]{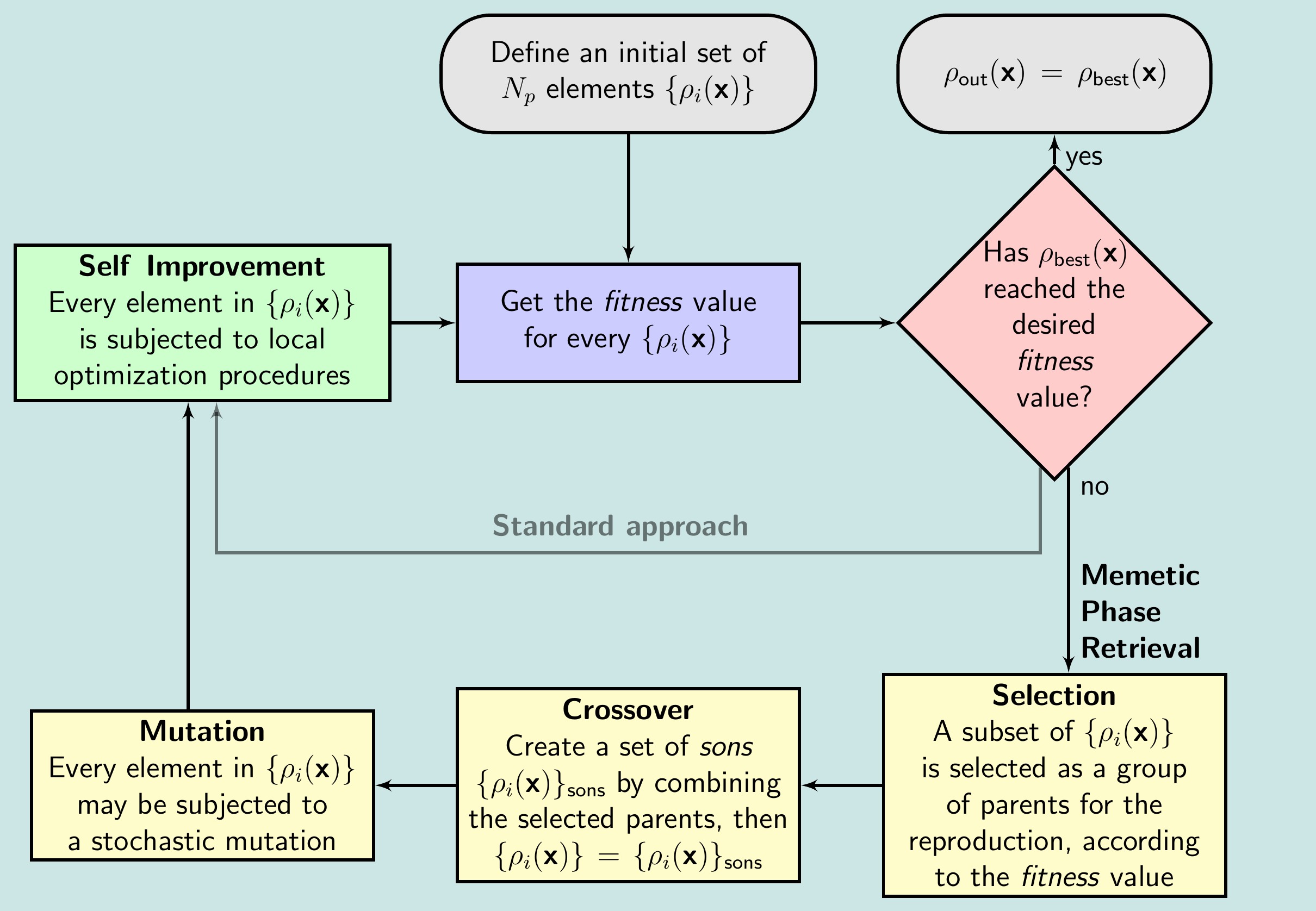}
\caption{Comparison between the MPR algorithm and the standard phasing approach.}
\label{MPR_diag}
\end{figure}

Using this method, we introduce local iterative deterministic phase retrieval procedures in our algorithm as \emph{Self Improvement} operation.
For this reason, the new proposed phasing approach is called Memetic Phase Retrieval (MPR). Fig.\ref{MPR_diag} shows an overview on the procedure, while a more detailed description of the method is reported in the Methods section. It is worth noting that the standard phasing approach can be seen as MPR without the genetic operators \emph{Selection}, \emph{Crossover} and \emph{Mutation}. 
MPR has to be considered as a smart framework to take advantage and improve the performances of any current iterative phase retrieval approach; it is clear, in fact, that any phase retrieval algorithm can be implemented in MPR.
In this work we use the Error Reduction and Hybrid Input-Output algorithms as \emph{Self Improvement} operations because they are simple, well known and well-characterized methods. Moreover, as we will show in the following, the sole inclusion of HIO and ER inside MPR is enough to build a very powerful phase retrieval algorithm. 
MPR actually includes also methods for the retrieval of the optimal support function, like the \emph{Shrinkwrap} algorithm \cite{Marchesini2003}; this feature makes MPR a Co-evolving Memetic Algorithm \cite{Smith2007}, where the \emph{Self Improvement} co-evolves along with candidate solutions. This peculiar feature will be discussed in future works because it has not been used in the present application on KEDI data, where sufficient information about the support function was available, making \emph{Shrinkwrap} procedure unnecessary.


\subsection*{Testing MPR on simulated data}
In the Supplementary Information we discuss in detail several numerical tests performed to verify the potentialities of the new proposed approach to reliably retrieve in a reliable way phase information in comparison to standard deterministic phasing approaches. Here, we reassume the main results obtained by simulations before discussing the application of MPR on true experimental data of a KEDI experiment.

In the comparison of the performances between MPR and standard phasing algorithms, applied on simulated data, we will focus on the best phase reconstructed by both methods and not on a statistical analysis of the results obtained from the set of phase retrieval processes started in parallel with different initial conditions. This is necessary because the genetic dynamics has the distinctive feature to mix and share parallel information during the stochastic evolution; this makes smarter the stochastic search for a better phase, but also tends to push the population near to the best candidate solution.
A statistical analysis would be thus biased in favor of MPR. 
Moreover, we are going to compare MPR and standard phasing algorithms under equal conditions: the same number of candidate solutions in the population and the same kind of iterative phase retrieval algorithms.

\begin{figure}
  \includegraphics[width=\linewidth]{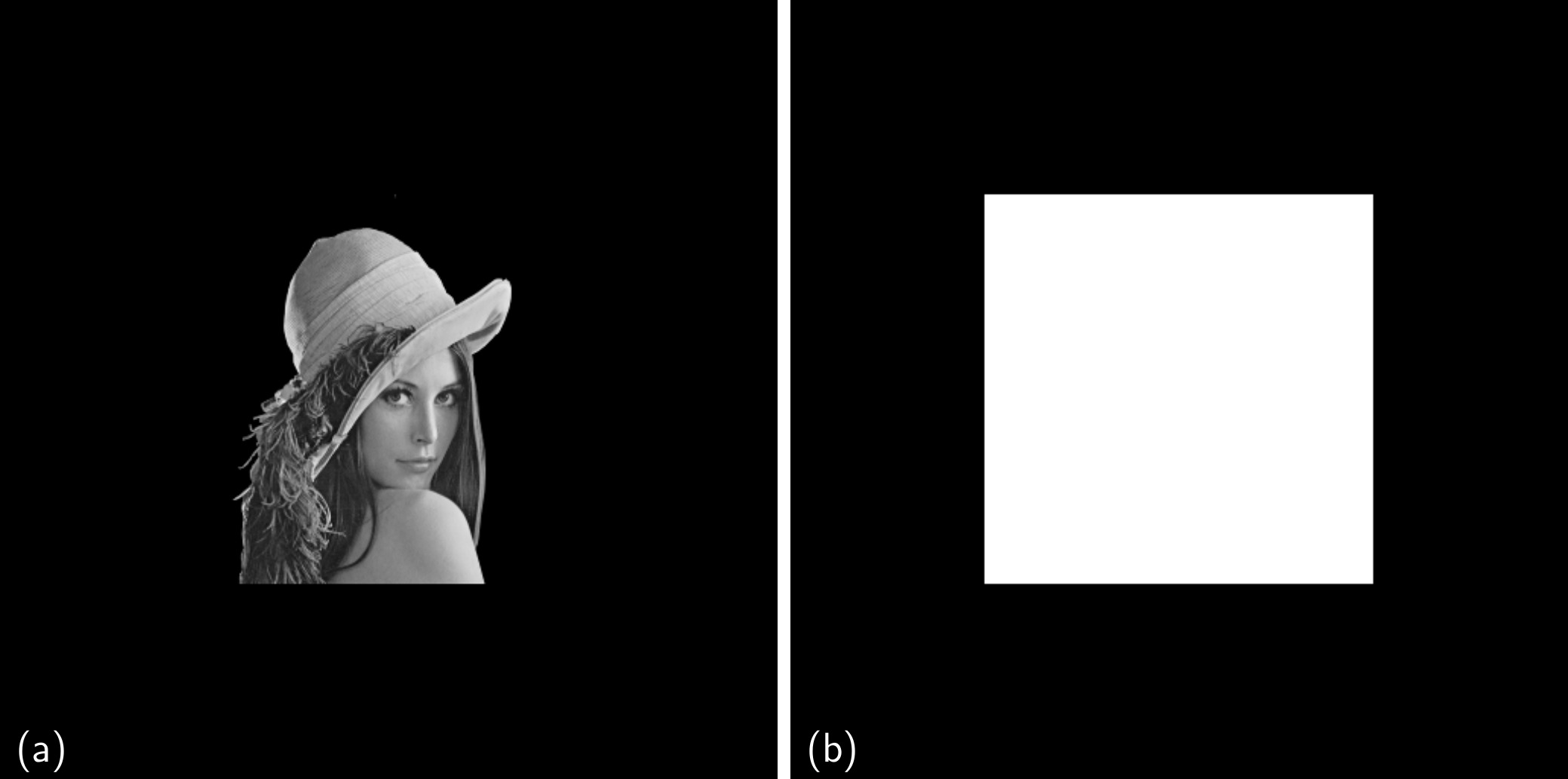}
\caption{Real-valued data. (a) Unknown test function (Lena, adapted from the picture 4.2.04 in the USC--SIPI image database\cite{SIPI1997}), (b) support function.}
\label{fig:real_input}
\end{figure}

The first test of comparison between MPR and the standard phasing approach is a phase retrieval of a positive and real-valued two-dimensional (2D) scattering function, as shown in Fig.\ref{fig:real_input}a.
In the test the provided support function \emph{S} (Fig.\ref{fig:real_input}b) is a square four times smaller than the total area of the direct space, which gives a \emph{constraint ratio} \cite{Millane2015} $\Omega=2$.
With this regard it is useful to remember that $\Omega \geq 1$ is the mathematical condition to assure the existence of a unique solution to the phase problem. $\Omega$ is defined as the ratio between the support of the autocorrelation of the scattering object and two times the object support, which are areas for 2D phase problem, volumes for 3D ones. In this example, the support function is not updated during the phase retrieval.

\begin{figure}
 \centering
  \includegraphics[width=\linewidth]{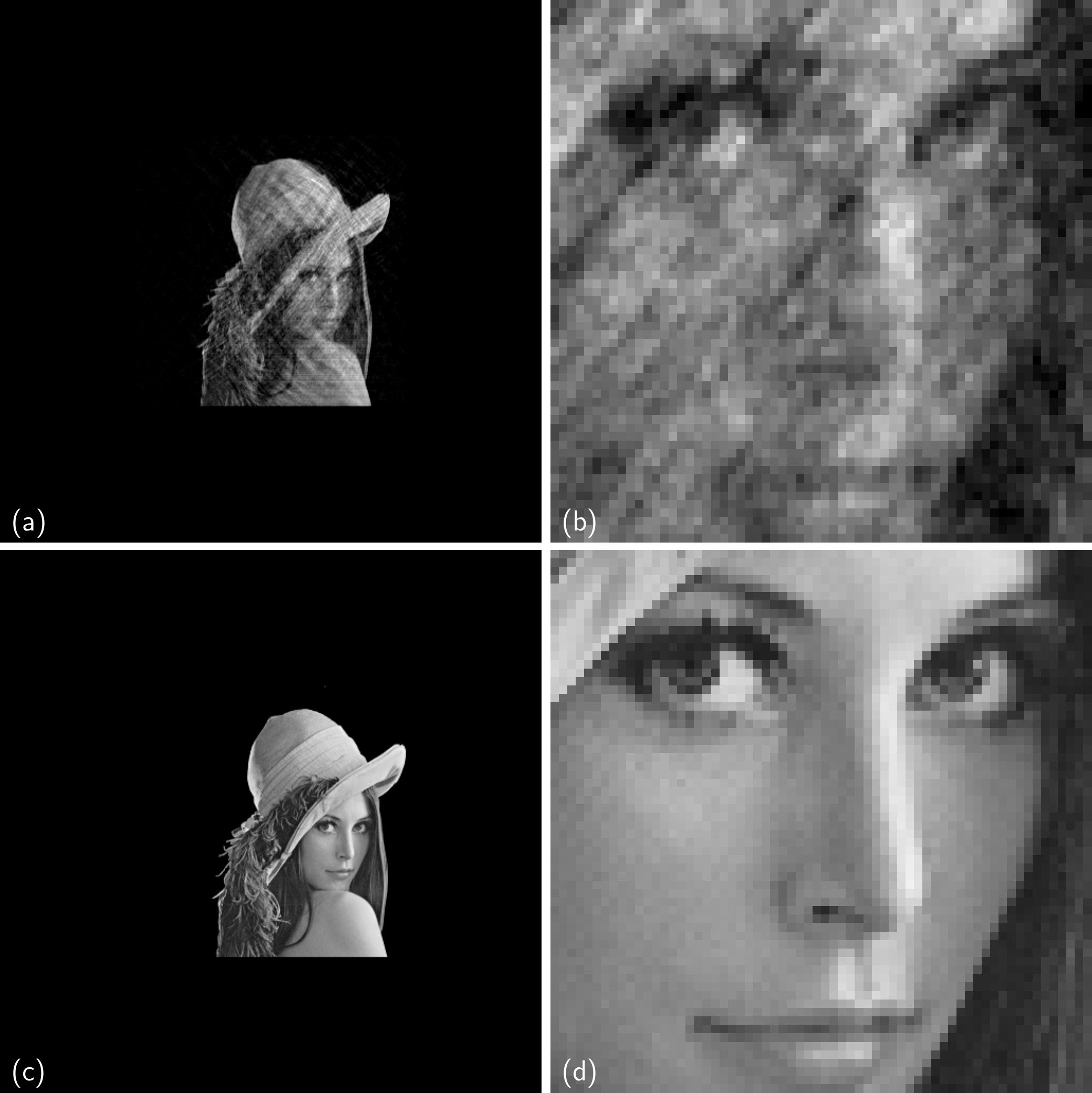}

\caption{Real-valued phasing test. Retrieved function (a) and a zoom (b) for the standard approach; retrieved function (c) and a zoom (d) for MPR.}
\label{fig:real_comp}
\end{figure}

Fig.\ref{fig:real_comp} shows the obtained results for a population $N_p = 16384$. In particular, Fig.\ref{fig:real_input}b shows the support \emph{S}, never updated during the phasing process. Fig.\ref{fig:real_comp}a shows the retrieved unknown function obtained by standard phasing algorithms (see Supplementary Information for further details). The poor quality of the retrieved phase is due to the weak constraint in real space, as the support constraint \emph{S} has not been updated. Fig.\ref{fig:real_comp}c shows the retrieved unknown function obtained by MPR (see \emph{Phase retrieval of real-valued data} in Supplementary Information for further details).

The performances of the new proposed stochastic approach are much better than the classic deterministic phasing methods. MPR works accurately even without a tight support constraint.

In the case of simulated data one can evaluate the true error defined as the normalized absolute difference for every pixel of the retrieved 2D function with respect to the solution (Lena image, adapted from the picture 4.2.04 in the USC--SIPI image database\cite{SIPI1997}) (see Supplementary information for more details).
The true error for the deterministic phasing approach is larger than $24\%$. Instead, MPR leads to a true error less than $1\%$.

\begin{figure}
  \centering
  \includegraphics[width=\linewidth]{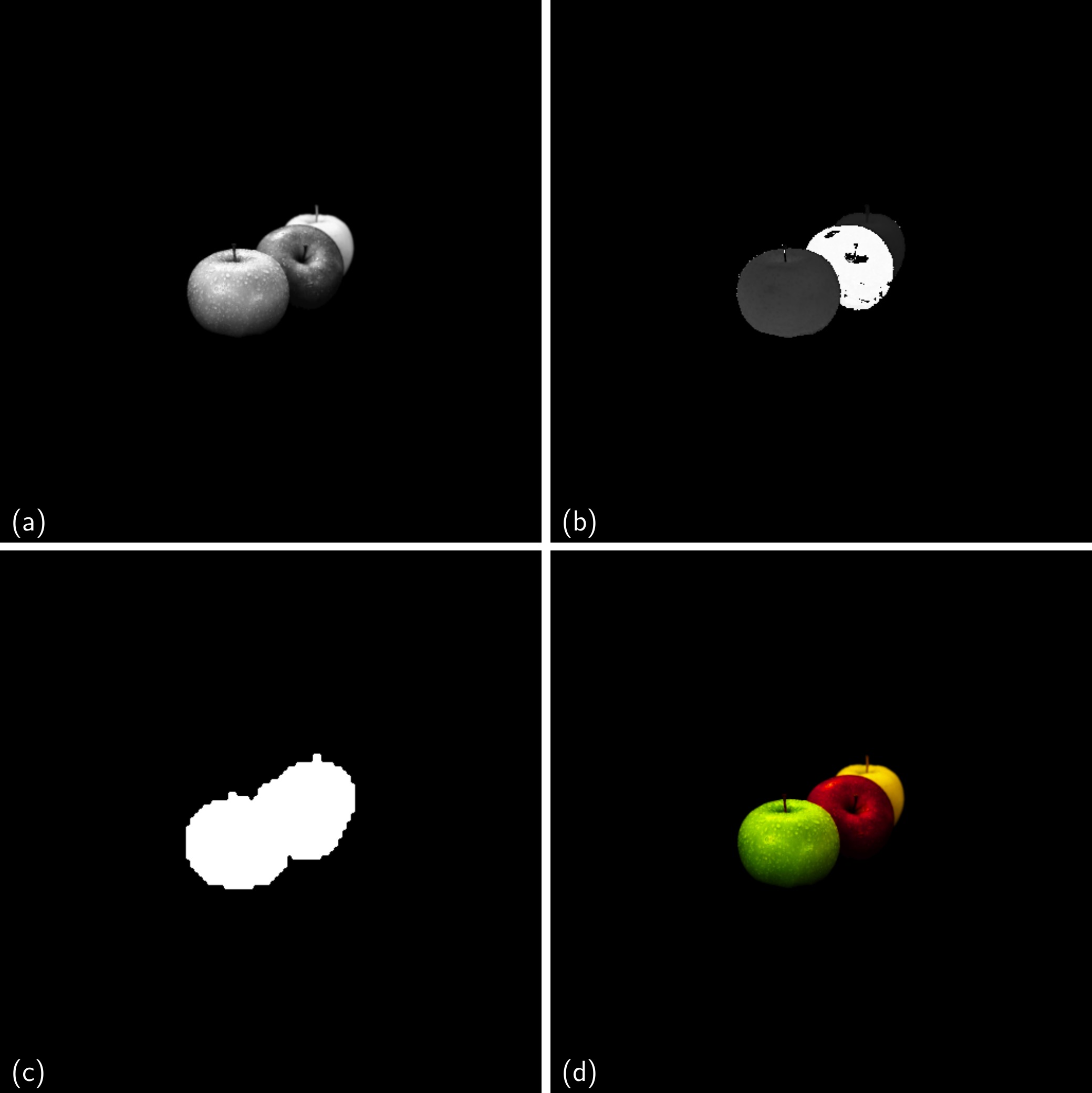}

\caption{Complex-valued data. (a) Unknown modulus of the complex-valued function (modulus of the solution); (b) unknown phase of the complex-valued function (phase of the solution), with range 0-360$^{\circ}$ ; (c) initial support function; (d) representation of the complex solution (adapted from the image at \textit{http://openwalls.com/image?id=17210}, Copyright Creative Commons Attribution 3.0 Unported), assigning to the brightness the values of the modulus and to the hue the values of phases, following the Hue-Saturation-Value (HSV) color system.}
\label{fig:complex_data}
\end{figure}

A second numerical test of MPR concerns the reconstruction of a complex-valued scattering function.
In particular, we have considered a situation typically encountered in EDI, in which HRTEM allows us to obtain a lens-based image of the sample under investigation characterized by a worse spatial resolution with respect to that corresponding to the $\text{NA}_\text{diff}$ of the measured diffraction pattern. In order to simulate this experimental situation, we have binned with a factor 4 the module of the scattering function to be retrieved (Fig.\ref{fig:complex_data}a) to obtain a rough estimation of its support, thresholding the binned image as shown in Fig.\ref{fig:complex_data}c, whereas Fig.\ref{fig:complex_data}b shows the phase in direct space that has also to be retrieved. 

\begin{figure}

  \includegraphics[width=\linewidth]{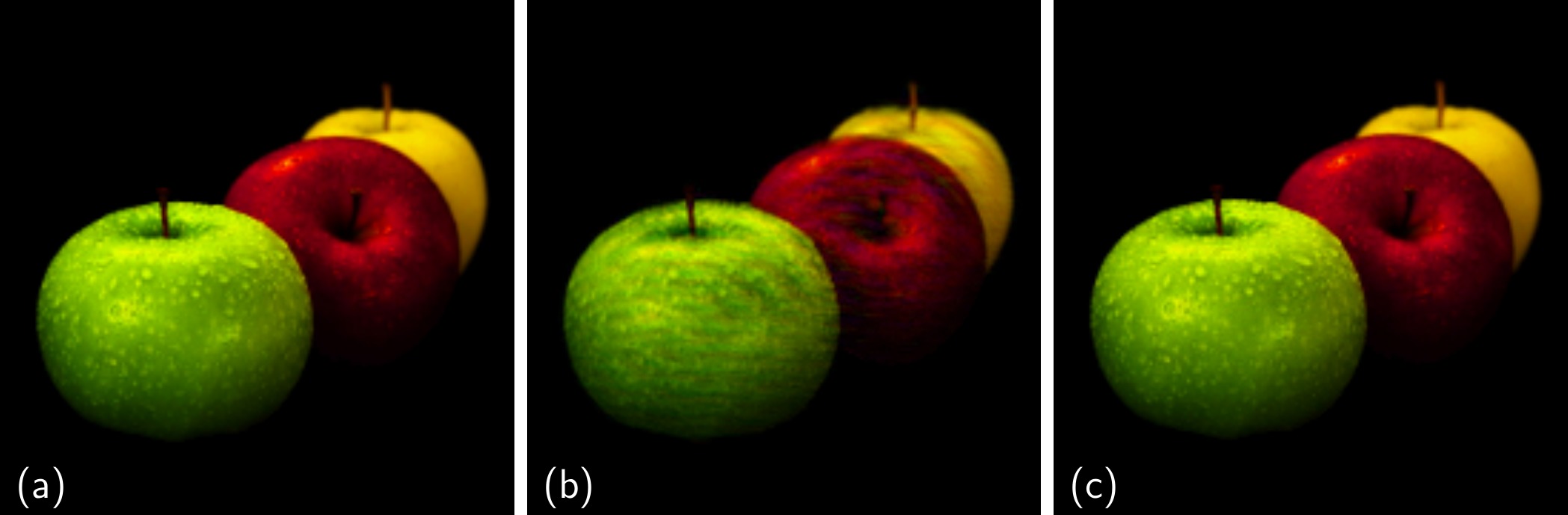}

\caption{ Complex-valued phasing test. (a) Detail of the solution (adapted from the image at \textit{http://openwalls.com/image?id=17210}, Copyright Creative Commons Attribution 3.0 Unported); (b) Retrieved scattering function for the standard approach; (c) Retrieved scattering function for MPR. The values of the modulus are assigned to the brightness while the values of phases are assigned to the hue, following the Hue-Saturation-Value (HSV) color system.}
\label{fig:complex_comp}
\end{figure}

This data can be represented as depicted in Fig.\ref{fig:complex_comp}a via the Hue-Saturation-Value (HSV) color system, where the information on the phase is stored in the hue, the modulus corresponds to the value and the saturation level is set to the maximum.
Even in this case, the standard phasing approach is far from recovering the correct complex scattering function, reported in Fig.\ref{fig:complex_comp}b. Instead, MPR is able to correctly retrieve both module and phase of the complex unknown scattering function (Fig.\ref{fig:complex_comp}c) (see Supplementary Information for further details).
The final true error for this test is about $10\%$ for the standard approach and $1.5\%$ for MPR.

\subsection*{Application of MPR on experimental data for Keyhole Electron Diffractive Imaging}

KEDI experiments \cite{DeCaro2012} are challenging tests for phase retrieval algorithms as the scattering function to be reconstructed is complex. $\text{SrTiO}_\text{3}$ was considered as a case study for the great importance of this oxide from both an applicative and a fundamental point of view. The role of the oxygen sub-lattice is of particular importance in the studies of two dimensional electron gases formed at the interface between two insulating oxides and has recently attracted great attention \cite{Banerjee2015}. The capability to image the lattice of complex oxides at atomic resolution is necessary to understand the intriguing properties of this class of material. Moreover imaging of a light chemical element, such as oxygen, in a matrix of heavier atoms, like titanium and strontium, is not straightforward \cite{Varela2006, Muller2004}. The samples were prepared for KEDI experiments in a [100] zone axis as this configuration enables the imaging of different atomic species in the crystal sub-lattice (see Methods section).


\begin{figure}

  \includegraphics[width=\linewidth]{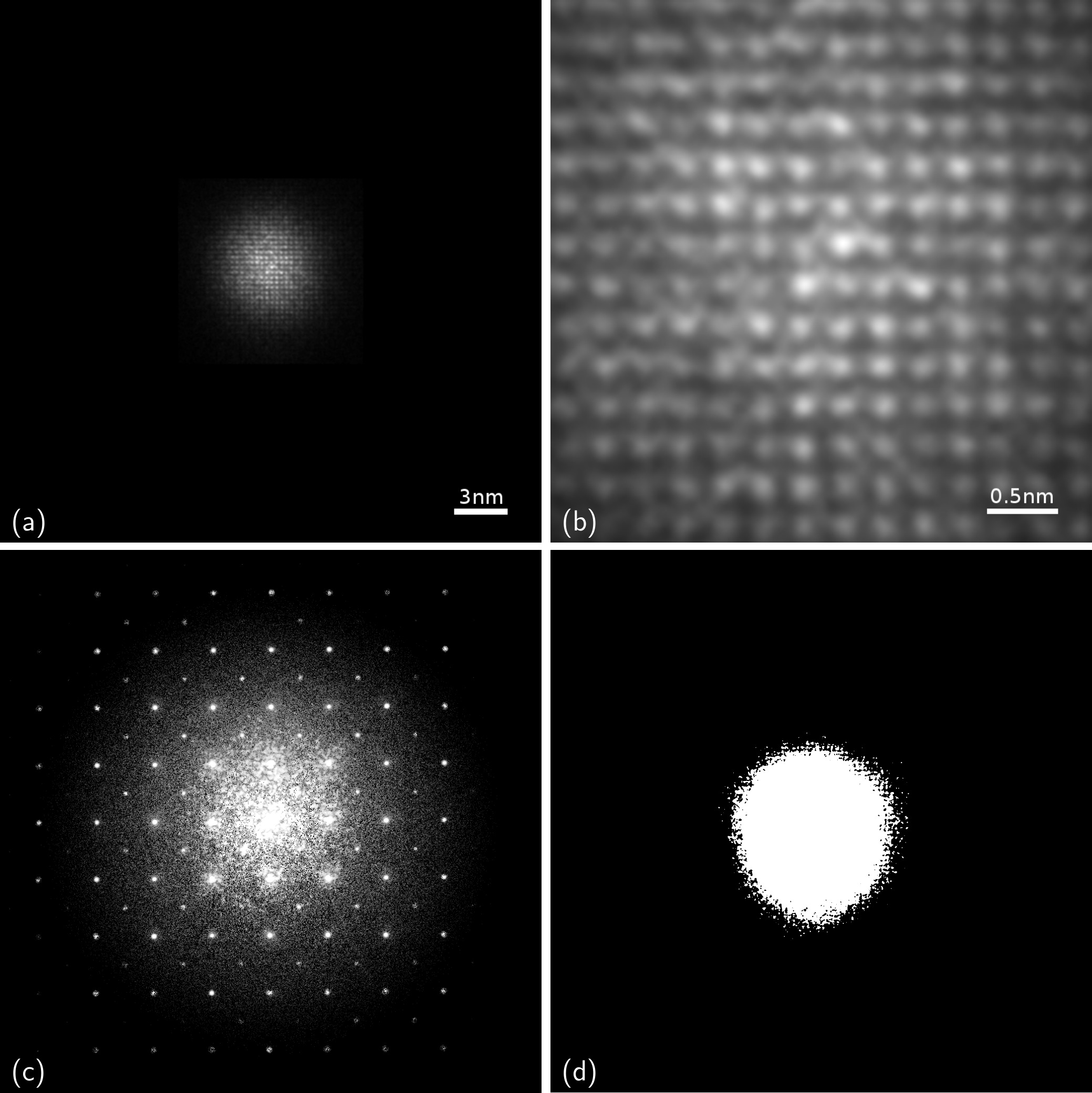}

\caption{KEDI data. (a) Experimental HRTEM image in [100] zone axis and (b) zoom; (c) KEDI diffraction pattern used as input for the phasing; (d) initial support function extracted from (a).}
\label{fig:input_STO}
\end{figure}

KEDI requires an HRTEM image (Fig.\ref{fig:input_STO}a) and a nano-diffraction pattern acquired from the same sample area with the same electron optical conditions \cite{DeCaro2012}.
The HRTEM experiment enables us to image the sample (Fig.\ref{fig:input_STO}a and \ref{fig:input_STO}b), to complement the diffraction pattern at the lower spatial frequency (Fig.\ref{fig:input_STO}c) and to estimate the support (Fig.\ref{fig:input_STO}d) at the resolution allowed by the experimental conditions and by the electron objective lens aberrations \cite{DeCaro2012}. In the case of the electron-optical set-up used for these experiments the relevant spatial resolution in the HRTEM image at optimum defocus is 0.19 nm \cite{Carlino2014}. The HRTEM image in Fig.\ref{fig:input_STO}b has been successfully simulated in the framework of full dynamical Bloch-wave approach \cite{Spence2013} for a thickness of 25 nm and an underfocus value of 41.3 nm (see Supplementary Information). It should be noted that the phase contrast in the HRTEM image of Fig.\ref{fig:input_STO}b does not show any evident clues that could be correlated to the presence of the oxygen atomic columns which should be seen in the [100] projection of the $\text{SrTiO}_\text{3}$ atomic potential, as evidenced in the simulation (see Fig. S7 of Supplementary Information).
Indeed, it is worthwhile to remark that the HRTEM image is, in general, an interference pattern of the waves scattered by the atomic potentials in the specimen, and the positions of the maxima and minima in the image cannot be straightforwardly interpreted as structural features and therefore the comparison with the simulated images is needed \cite{Spence1988} (see Fig. S7 of Supplementary Information).

The KEDI diffraction pattern, shown in Fig.\ref{fig:input_STO}c has been obtained by combining the measured diffraction pattern with the modulus of the HRTEM image FT, after a suitable matching procedure requiring its rotation and scaling \cite{DeCaro2012}. 
The pattern in Fig.\ref{fig:input_STO}c is the starting point for the phase retrieval process. 
It is worth noting that the MPR phasing process has been carried out without any a--priori information about the phases, an information which is, instead, needed by standard phasing procedures applied to KEDI \cite{DeCaro2010}.
Further details on MPR applied to experimental data have been reported in the Supplementary Information.

\begin{figure}
  \centering
  \includegraphics[width=1\linewidth]{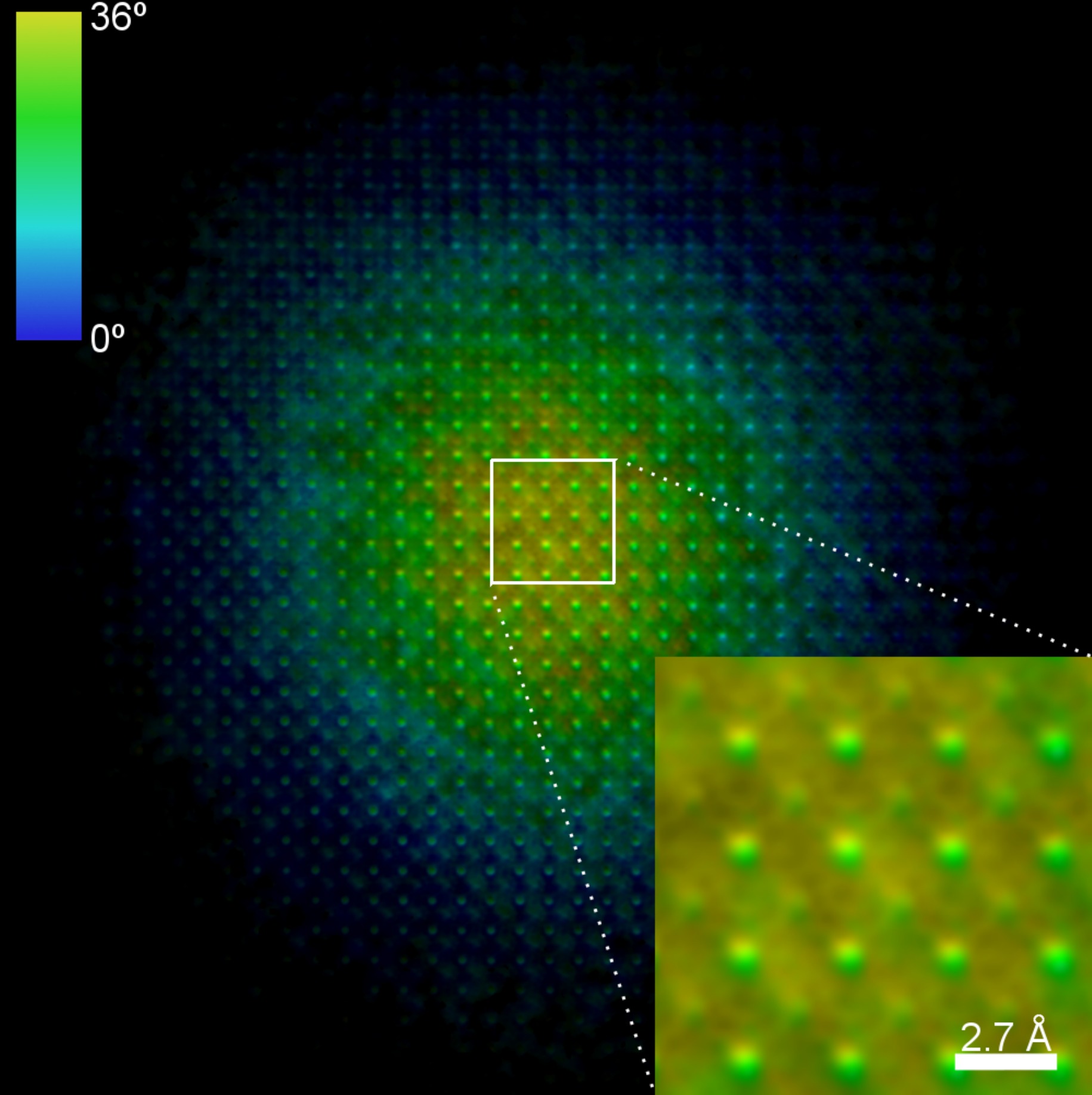}
  \caption{Recovered image. Brightness is proportional to the retrieved modulus and colors indicate the retrieved phase.}
  \label{fig:STO_colored}
\end{figure}

Fig.\ref{fig:STO_colored} shows the retrieved scattering function obtained by using MPR, where the brightness corresponds to the modulus and the hue to the phase of the retrieved real-space complex-valued scattering function. 
The long range phase variation is due to the phase variation of the illumination nano-probe \cite{DeCaro2016}.

\begin{figure}

  \includegraphics[width=1\linewidth]{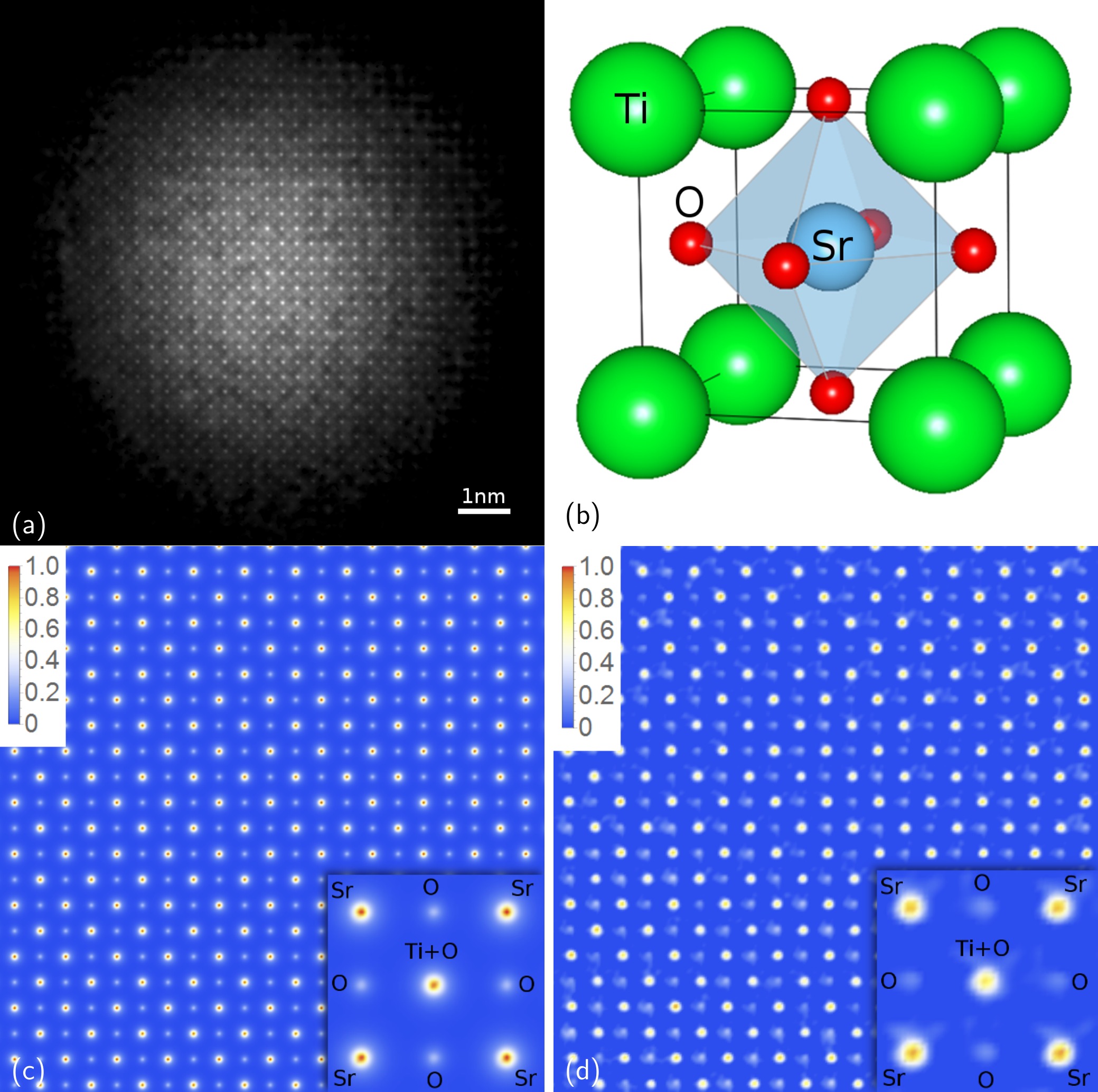}

\caption{Results of MPR on experimental KEDI data. (a) Modulus of the retrieved scattering function (relevant to \ref{fig:input_STO}a); (b) $\text{SrTiO}_\text{3}$ unit cell; (c) simulation of the $\text{SrTiO}_\text{3}$ projected potential in [100] zone axis and (d) experimental data extracted from the phased map.}
\label{fig:STO_res}
\end{figure}

Fig.\ref{fig:STO_res}a shows the phase retrieved amplitude for the structure of the $\text{SrTiO}_\text{3}$ seen in a [100] projection.
Fig.\ref{fig:STO_res}d has been obtained by subtracting the contribution of the TEM illumination function to quantify the $\text{SrTiO}_\text{3}$ projected potential.
The first important point that should be emphasised, is that in the phase recovered image the positions of the maxima are correctly in correspondence with the expected positions of the atomic columns seen in the [100] projection. In other words, the phase reconstructed image is a structural image of the specimen. This is a fundamental issue that paves the way for quantitative structural imaging at atomic resolution. 
Indeed, as shown in Fig.\ref{fig:STO_res}d, the Sr and Ti+O columns are precisely seated on the relevant square sublattice of the $\text{SrTiO}_\text{3}$ in the [100] projection (see Fig.\ref{fig:STO_res}c and Fig.\ref{fig:STO_res}b). Approximately in the center of the sublattice there is a lower signal which corresponds to the oxygen columns.
A second point concerns the retrieval of quantitative information about the atomic potential. By comparing data shown in Fig.\ref{fig:STO_res}d with the expected projected atomic potential (Fig.\ref{fig:STO_res}c) we found that the ratios between the intensities of Sr, Ti+O and O columns are correctly retrieved, providing truly quantitative information on the specimen. In particular, the expected intensity ratios are $\text{I}_\text{O} / \text{I}_\text{Sr}=0.35$ and $\text{I}_\text{Ti+O} / \text{I}_\text{Sr}=0.96$  while the experimental retrieved data give $\text{I}_\text{O} / \text{I}_\text{Sr}=0.35\pm0.05$ and $\text{I}_\text{Ti+O} / \text{I}_\text{Sr}=0.89\pm0.10$.
The possibility to do KEDI experiments by MPR reconstruction paves the way for a detailed structural characterization of the investigated samples and opens up new possibilities for the understanding of the properties of the matter at sub-\r{A}ngstr\"om resolution.

\subsection*{Conclusions}
In this work we have discussed some of the potentialities of a new phasing approach, stochastic in the exploration of the space of solutions, based on a memetic algorithm, applicable both to X-ray and electron coherent diffraction imaging. We have tested the new phasing algorithm, named Memetic Phase Retrieval, on simulated data and we have obtained the result that the knowledge of the support - which defines the boundaries in the direct space of the unknown scattering function that one wants to retrieve - is less binding than previously reported. The more efficient exploration of the space of solution, possible thanks to the stochastic genetic procedures implemented in MPR, should be the higher gear of the new proposed phasing method with respect to those already available. Indeed, both the possibility to correctly retrieve a real-valued scattering function by its diffraction pattern without imposing any tight support and to reconstruct a complex-valued scattering function by using a low-resolution estimate of the support, are examples of the great capabilities of MPR to face the phase problem. Our tests on simulated data demonstrate the superior capabilities of the MPR for accurate phase retrievals. Indeed, by using the same computational resources, the MPR approach has proved to be much more powerful than deterministic phasing procedures in facing the phase problem. 
The application to an experimental case of Keyhole Electron Diffraction Imaging has shown that the atomic potentials of $\text{SrTiO}_\text{3}$ can be quantitatively imaged, representing a relevant improvement for the study of the matter.
We believe that the Memetic Phase Retrieval approach could be of interest in all the fields that require accurate phase retrievals.

\section*{Methods}

\subsection*{The phase problem as an \emph{optimization problem} }\label{phaseopt}
The ideal solution to the phase problem is a function $\rho_s (\vec{x})$, representing the spatial distribution of the sample, whose Fourier Transform (FT), $\tilde \rho_s (\vec{q})$, has a square modulus equal to $I(\vec{q})$, which is proportional to the experimental diffraction pattern intensity. 
$\rho_s (\vec{x})$ is also assumed to be zero outside a well-defined region of the real space, the so-called support \emph{S}, in order to satisfy the oversampling condition, which assures the necessary information to retrieve $\rho_s (\vec{x})$ \cite{Sayre1952}. 
Here, $\vec{x}=(x,y)$, with $x$ and $y$ the cartesian components of the position vector x with respect to the reference system. 
Analogously, $\vec{q}=(u,v)$, where $u$ and $v$ are the spatial frequencies components with respect to the reference axes.

In principle, this solution can be represented as an intersection of sets.
$\mathcal{M}$ can be defined as the set of all functions $\rho (\vec{x})$ compatible with experimental data $I(\vec{q})$, i.e:
\begin{equation} \label{setm}
 \mathcal{M} = \{ \rho (\vec{x}) : | \tilde{\rho} (\vec{q}) |^2 = I(\vec{q}) \}.
\end{equation}

$\mathcal{S}$ is, instead, the set of all functions $\rho_s (\vec{x})$ satisfying the oversampling condition, which is defined by a binary function $\Pi (\vec{x})$ representing the object support \emph{S}. So, the set $\mathcal{S}$ is described by:

\begin{equation} \label{sets}
 \mathcal{S} = \{ \rho (\vec{x}) : \rho(\vec{x})= \Pi (\vec{x}) \rho(\vec{x}) \}.
\end{equation}

Thanks to \eqref{setm} and \eqref{sets}, the ideal solution is:
\begin{equation} \label{idsol}
 \rho_s (\vec{x}) = \mathcal{M} \cap \mathcal{S}.
\end{equation}

The main issue concerning experimental measurements is the presence of noise and lack of data; this, in general, implies that $\mathcal{S}$ and $\mathcal{M}$ do not intersect:
 \begin{equation} \label{nosol}
 \mathcal{M} \cap \mathcal{S} = \varnothing.
\end{equation}

Due to the condition \eqref{nosol} a different way to define what we mean by ``solution'' is needed. 
It is useful, at this point, to introduce two projection operators, which act on the function $\rho (\vec{x})$:

\begin{equation} \label{pm}
P_\mathcal{M} : P_\mathcal{M} \rho(\vec{x}) = \mathscr{F}^{-1} [ \sqrt{I(\vec{q})} e^{i \arg{[\tilde{
\rho} (\vec{q})}]}](\vec{x}),
\end{equation}

\begin{equation} \label{ps}
P_\mathcal{S} : P_\mathcal{S} \rho(\vec{x}) = \Pi(\vec{x}) \rho(\vec{x}).
\end{equation}

It's trivial to prove that $P_\mathcal{M}$ and $P_\mathcal{S}$ are projectors on sets $\mathcal{M}$ and $\mathcal{S}$, previously defined in \eqref{setm} and \eqref{sets}.
Thanks to these operators, it's now possible to give a new definition of solution in place of the one defined in the eq. \eqref{idsol}:
\begin{equation} \label{realsol}
\rho_s (\vec{x}) = \min_{\substack{\rho}} D[P_\mathcal{M} \rho(\vec{x}), P_\mathcal{S} \rho(\vec{x})],
\end{equation}
where the functional $D[A,B]$ represents the metric of the space. Hereafter, we will refer to the eq. \eqref{realsol} whenever we will talk about the ``solution'' of the problem, $\rho_s (\vec{x})$.

It's now clear that, in this framework, finding a solution to the phase problem means minimizing the distance between sets $\mathcal{M}$ and $\mathcal{S}$: the \emph{phase problem} becomes an \emph{optimization problem} for the quantity $D[P_\mathcal{M} \rho(\vec{x}), P_\mathcal{S} \rho(\vec{x})]$, which can be reinterpreted as the \emph{error} of the recovered density $\rho(\vec{x})$.
Different definitions of the metric imply different definitions of the \emph{error} assigned to a given $\rho(\vec{x})$ and, as consequence, different optimization targets.
We can define the \emph{error functional} $E[\rho]$ as
\begin{equation} \label{errfunc}
E[\rho] = D[P_\mathcal{M} \rho(\vec{x}), P_\mathcal{S} \rho(\vec{x})],
\end{equation}
such that the eq. \eqref{realsol} turns into
\begin{equation} \label{realsol_errfunc}
\rho_s (\vec{x}) = \min_{\substack{\rho}} E[\rho].
\end{equation}

Standard approaches to the phase problem are mainly deterministic iterative algorithms which, going back and forth from the real to the Fourier space, try to minimize a specific \emph{error functional} \cite{Fienup1982}. These methods are highly efficient in finding local minima, but they suffer from stagnation and the final result is highly dependent on the initial conditions \cite{Marchesini2007}.
In order to overcome these issues, phasing procedures usually involve a lot of parallel and independent retrieval processes with different initial conditions and then selecting the one with the lowest error.

The founding idea of the new proposed phasing method is to better perform this parallel exploration of the space, through the use of a Memetic Algorithm.

\subsection*{Selection as a Rigged Roulette}
The Selection process is a delicate step in the Evolution process. A Selection strongly favoring only the better elements in $ \{ \rho_i(\vec{x}) \}_{i=1 \ldots N_p}$ (i.e., elements with the better \emph{fitness} value) will improve the convergence speed, but the algorithm will suffer with stagnation in local minima.
On the other side, a selection process that weakly favors those elements will have, instead, an unstable convergence and will require an excessive length of time to find the solution.

There are several ways to select elements depending on their \emph{fitness} value. The one chosen in this work is the so-called ``rigged roulette''.
Once an error value $E_i$ is assigned to every $\rho_i(\vec{x})$ in $ \{ \rho_i(\vec{x}) \}_{i=1 \ldots N_p}$ according to the eq.\eqref{errorer}, the set $ \{ \rho_i(\vec{x}) \}_{i=1 \ldots N_p}$ is ordered by increasing values of $E_i$ (which is equivalent to a decreasing values of the \emph{fitness}).
Whenever the algorithm has to select a $\rho_i(\vec{x})$ in $ \{ \rho_i(\vec{x}) \}_{i=1 \ldots N_p}$ for the Crossover operation, an index is extracted through the relation
\begin{equation}\label{riggedroulette}
s = \lfloor \{ \text{rand}[0,1)\}^r \cdot N_p \rfloor + 1 ,
\end{equation}
where $r \geq 1$ is related to the ``strength'' of the selection process. Usual values of $r$ range from $1.5$ to $2.5$.

Eq. \eqref{riggedroulette} maps a flat distribution in $ [1, 0) \subset \mathbb{R}$ to an unbalanced distribution in $ \{ 1, N_p \} \subset \mathbb{Z} $, where the higher the value is the greater the probability is of getting a lower index and, therefore, selecting a better element.

\subsection*{Differential Crossover}
In the Natural Evolution process, the Crossover operation is the mixing of the parents' genetic pool.
In our implementation chromosomes are represented by every single (complex) value of $\tilde{\rho} (\vec{q}) = \mathscr{F}[\rho] (\vec{q})$.
This means that, given two parent functions $\rho_1(\vec{x})$ and $\rho_2(\vec{x})$ selected according to their fitness, the son function $\rho_\text{son}(\vec{x})$ is created according to

\begin{equation} \label{cross}
\tilde{\rho}_\text{son}(\vec{q}) = 
\begin{cases} 
\tilde{\rho}_1 (\vec{q}), & \mbox{if rand}[0,1) > C \\ 
\tilde{\rho}_2 (\vec{q}), & \mbox{otherwise}, 
\end{cases}
\end{equation}
where rand$[0,1)$ is a random number with flat distribution in $[0,1)$ and C is a balancing coefficient between 0 and 1.

An improvement in performances can be obtained using the so called Differential Crossover \cite{Storn1997} where, instead of selecting two parents, four parents, $\rho_1(\vec{x})$ $\rho_2(\vec{x})$ $\rho_3(\vec{x})$ and $\rho_4(\vec{x})$, are chosen.
The differential crossover acts as follows:

\begin{equation} \label{crossdiff}
\tilde{\rho}_\text{son}(\vec{q}) = 
\begin{cases} 
\tilde{\rho}_1 (\vec{q}), & \mbox{if rand}[0,1) > C \\ 
\tilde{\rho}_2 (\vec{q}) + \\ \quad D_c \cdot [\tilde{\rho}_3 (\vec{q})-\tilde{\rho}_4 (\vec{q})] & \mbox{otherwise}, 
\end{cases}
\end{equation}
where $D_c$ is called \emph{differential coefficient} with typical values between $0.5$ and $1.5$.

The population of sons can be, in general, smaller than the whole population. This means that if the population of sons has $N_s = G \cdot N_p$ elements, a fraction of $N_p - N_s$ parents, chosen randomly, will survive to the next generation.

The parameter $G$, which can be called \emph{genetic fraction}, has values between $0$ and $1$: $G=1$ means that all of the parent population is replaced by the sons, while $G=0$ means that no sons are created, the genetic operators are switched off and we get a situation equivalent to the standard deterministic approach, as depicted in Fig.\ref{MPR_diag}. 

\subsection*{Mutation}

Every element in the population $ \{ \rho_i(\vec{x}) \}_{i=1 \ldots N_p}$ may be subjected to a stochastic modification.
In this work, the mutation operation has been switched off because it does not introduce a remarkable improvement in the performance of MPR on treated data. Different implementations of the mutation operator are under study and will be topics of future works.

\subsection*{Self improvement via deterministic optimization} 
Optimization algorithms such as Error Reduction (ER) and Hybrid Input-Output (HIO) are efficient methods to find local minima or, more precisely, minima bounded to a region of the configuration space near the starting point. 
These algorithms are strictly bounded to the metric $D[A(\vec{x}), B(\vec{x})]$ defined as:
\begin{equation}\label{metrer}
D[A(\vec{x}), B(\vec{x})] = \int d\vec{q} \; [\; |\mathscr{F}[A](\vec{q})| -  |\mathscr{F}[B](\vec{q})|\;]^2.
\end{equation}

This implies that the local optimization target is the functional $E[\rho]$ defined as:
\begin{multline}\label{errorer}
E[\rho] = D[P_\mathcal{M} \rho(\vec{x}), P_\mathcal{S} \rho(\vec{x})] = \\ = \int d\vec{q} \; [ \sqrt{I(\vec{q})}  -  |\mathscr{F}[\Pi \rho](\vec{q})|\;]^2.
\end{multline}

In our algorithm this local optimization is carried on by elaborating every $\rho_i(\vec{x})$ with $N_\text{HIO}$ iterations of the Hybrid Input-Output algorithm and $N_\text{ER}$ iterations of the Error Reduction algorithm. 
In this work the global optimization target, i.e., the \emph{fitness} of MPR, coincides with the local optimization target of ER and HIO algorithms just shown in \eqref{errorer}. 
This is not to be taken for granted because, in general, we can define any arbitrary global optimization target different from the local one \eqref{metrer}.
We are testing different \emph{fitness} definitions for the global optimization, like the Csiszar's Information Divergence\cite{Csiszar1991},  and different local optimization algorithms.

\subsection*{The choice of the initial guess}
The phase retrieval process can be divided into two main steps. 
The first one concerns the choice of the initial \emph{population} of densities $ \{ \rho_i(\vec{x}) \}_{i=1 \ldots N_p}$. In the second step, we have to choose the parameters both of the genetic and the local optimization algorithms.

Standard approaches like Hybrid Input Output and Error Reduction need a single initial guess, which represent the first estimation of the solution.

MPR approach requires, instead, a set of initial guesses.
This set is produced from a single guess, simply randomly shifting every phase.
This means that, given an initial guess $\rho_\text{init} (\vec{x})$, every  $\rho_i(\vec{x})$ in $ \{ \rho_i(\vec{x}) \}_{i=1 \ldots N_p}$ is created via the relation
\begin{multline}\label{initguess}
\tilde{\rho}_i(\vec{q}_j) = \sqrt{I(\vec{q}_j)} \exp(\sqrt{-1}\phi_j) \text{ with }\phi_j = \\ =\arg[\tilde{\rho}_\text{init}(\vec{q}_j)] + R_c \cdot \text{rand}[-\pi, \pi].
\end{multline}

The parameter $R_c$, which has values between $0$ and $1$, depends on the accuracy of the initial guess $\rho_\text{init} (\vec{x})$.
If $\rho_\text{init} (\vec{x})$ is already a good estimation of the solution, it will be useful to set a low value (usually near to 0) for coefficient $R_c$ in order to well explore the space near $\rho_\text{init} (\vec{x})$.
If, instead, $\rho_\text{init} (\vec{x})$ is considered to be far from the solution, it is useful to set a value of $R_c$ near to $1$, in order to explore also areas of the space far from $\rho_\text{init} (\vec{x})$.

\subsection*{KEDI experiment}

A KEDI experiment was performed following the procedure reported in \cite{DeCaro2012} , which enables one to deliver a low dose of electrons to the specimen. The experiment requires the acquisition of an HRTEM image and a diffraction pattern by using the same electron optical set up. The experiments were performed by using a JEOL JEM 2010 F UHR operated at 200 kV. The cathode is a high coherence Shottky type. The microscope has an objective lens with low spherical aberration coefficient $C_s=0.47 \pm 0.01$ mm and a relevant resolution at optimum defocus in HRTEM of 190 pm. The environment around the microscope is thermally and mechanically very stable allowing us to achieve in the scanning TEM (STEM) high angle annular dark field (HAADF) mode a resolution of 126 pm, which is the theoretical limit for the used electron optical set up \cite{Carlino2005}. 

In a KEDI experiment the optical setup produces an electron nano-beam. The latter defines the mathematical support of the scattering function for the illuminated nanometric region of the extended crystal. As in a microscope the field of view is proportional to the inverse of magnification, the size of the illumination function (beam size) is somehow related to the spatial resolution. The electron beam size S (which defines the support) is directly related to the final resolution to be achieved and to the size of the detector used to record HRTEM image and n-ED pattern. In fact, if the highest frequency of the diffraction signal recorded in the reciprocal space is ${\rho^{-1}}$pm$^{-1}$, we should have $\rho$ at least two or three times the pixel size $\Delta_\text{map}$ of the phased map to have an electron projected potential two-dimensional map calculated with a sufficient number of points to be plotted continuously. 

For example, if we reached a final resolution – after the phase retrieval process – of $\rho=70$ pm we should have $\Delta_\text{map} \sim 25 \div 30$ pm which, multiplied by the detector pixel number along a line, N=1024 in our experimental case, would lead to a spatial region O (scattering region plus non-illuminated surrounding region) of $\sim 25 \div 30$ nm in size. Moreover, for the Nyquist theorem’s requirement, the illuminated beam size S (the support) has to be less than $2^{-\frac{1}{2}}$ O, i.e., at maximum $\sim 17 \div 20$ nm in size. Hence, in order to properly run the phase retrieval algorithms, the illuminated region of the sample in the direct HRTEM image has to be properly chosen with respect to the whole detector area to satisfy the above KEDI oversampling condition. 
Here, the cathode emission condition and the electron optical illumination system of the microscope has been experimentally set up to increase the probe coherence on the smallest illuminated area achievable \cite{DeCaro2012}. The microscope has an illumination system composed by three magneto-static lenses. These lenses were operated independently, together with the electrostatic lens of the emitter, to produce the smallest-sized probe on the focal plane of the pre-field of the objective lens and hence the smallest-sized coherent parallel beam on the specimen. The emission conditions of the microscope cathode were chosen to increase the coherence of the electron probe by decreasing the temperature of the emitting tip. We used a heating current for the filament that halves the emission current with respect to the standard operation, decreasing at the same time the electron dose delivered to the specimen. The current density on the specimen was below the detection limit of the amperometer connected to the phosphorus screen of the microscope ( $ < 0.1\; \text{pA} \; \text{cm}^{-2}$), allowing us to acquire the relevant diffraction pattern on the 1024x1024 Charge-Coupled Device (CCD) camera without using the beam stopper for the direct beam. Thus all the diffracted intensities were available for the phasing process and a very small dose is delivered to the specimen. The small electron probe, without any changes, was used to acquire both HRTEM image and diffraction from the same area of the specimen. Fig.\ref{fig:input_STO}a shows the HRTEM image. The illuminated area is $10\pm2$nm. The interference pattern of the phase contrast HRTEM image formed in the image plane of the objective lens is shown at a higher magnification in Fig.\ref{fig:input_STO}b In Fig.\ref{fig:input_STO}c the diffraction pattern formed in the back focal plane of the objective lens is shown. The central part of the pattern has been replaced, after proper scaling and rotation, by the FFT of the HRTEM image in Fig.\ref{fig:input_STO}a, as established in EDI method  \cite{Zuo2003,Huang2009,DeCaro2010}. The highest Miller’s index spot measurable in the pattern is the (5,5,0), which corresponds to a spacing of 55 pm. Thus, the expected gain in resolution of the maximum spatial frequency contained in the diffraction pattern ($\sim {\text{NA}_\text{diff}}^{-1}$) with respect to that corresponding to the FT of the HRTEM image is about four times.


\section*{Acknowledgments}
This work was supported by the NOXSS PRIN (2012Z3N9R9) project and Progetto premiale MIUR 2013 USCEF.
We acknowledge the CINECA and Regione Lombardia LISA award LI05p-PUMAS, 
for the availability of high-performance computing resources and support.

\section*{Author contributions statement}

A.C. and D.E.G. conceived the algorithm. A.C., supervised by D.E.G. and L.D., developed, implemented and tested the algorithm, and carried out the reconstructions. E.C. designed and performed the TEM EDI experiments, contributing to the optimization of the relevant data reduction and phasing. L.D. and F.S. made data reduction. All authors have equally contributed to the preparation and the revision of the text.

\section*{Additional Information}
\textbf{Competing financial interests:} The authors declare no competing financial interests.

\textbf{How to cite this article:} Colombo, A. et al. Facing the phase problem in Coherent Diffractive Imaging via Memetic Algorithms. Sci. Rep. 7, 42236; doi:10.1038/srep42236 (2017).


\end{document}